\documentclass[prb,a4paper,twocolumn]{revtex4}
\usepackage[dvips]{graphicx}

\begin{document}
\bibliographystyle{apsrev}


\title{Current suppression in a double-island single-electron transistor
for detection of degenerate charge configurations of a floating
double-dot}

\author{R.~Brenner}
\email[Electronic mail: ]{rolf@phys.unsw.edu.au}
\author{Andrew~D.~Greentree}
\author{A.~R.~Hamilton}
\affiliation{Centre for Quantum Computer Technology, School of
Physics, The University of New South Wales, Sydney NSW 2052,
Australia}

\date{\today}

\begin{abstract}
We have investigated a double-island single-electron transistor
(DISET) coupled to a floating metal double-dot (DD).
Low-temperature transport measurements were used to map out the
charge configurations of both the DISET and the DD. A suppression
of the current through the DISET was observed whenever the charge
configurations of the DISET and the DD were energetically
co-degenerate. This effect was used to distinguish between
degenerate and non-degenerate charge configurations of the
double-dot. We also show that this detection scheme reduces the
susceptibility of the DISET to interference from random charge
noise.
\end{abstract}

\maketitle

Single-electron transistors (SETs), the operation of which is
governed by Coulomb blockade effects, are highly sensitive
electrometers \cite{grabert1992}. Their capability to amplify
small charge signals on fast time scales \cite{schoelkopf98} has
made them interesting detectors for a variety of applications. In
particular, some alternative concepts to conventional computers -
namely quantum dot cellular automata (QCA) \cite{lent93} and
solid-state quantum computers (e.g.
Refs.~[\onlinecite{shnirman97,kane1998,vrijen00}]) - rely on
detection of small charge signals to read out a computational
result. The charge signals arise from different, spatial charge
distributions on two sites (such as metal islands, quantum dots or
donor atoms) that are separated by a tunnel barrier.

In this letter, we present work on a double-island SET (DISET, or
single-electron pump) \cite{pothier1992} capacitively coupled to a
floating double-dot (DD). We show that the DISET can sense
electrostatically degenerate charge configurations of the DD. This
capability arises from current suppression due to dynamic charge
correlations between the DISET and the DD, which was first
described in work related to QCA \cite{orlov1998,toth99}. We
suggest using this ability of the DISET for detecting the
propensity for charge motion between other spatially localised
sites, such as quantum dots and donor atoms. Uses for such a
detector could be in monitoring spin-charge conversion in
solid-state quantum computers, e.g. the models proposed by Kane
\cite{kane1998} and Vrijen \textit{et al.} \cite{vrijen00}.

The high charge sensitivity of SETs leads to high susceptibility
to random charge noise in the environment, which can result in
spurious signals. The operating conditions for the DISET are
different from those for conventional SET electrometers: whereas
the SET is generally biased to a point of maximum
transconductance, we operate the DISET at zero transconductance,
which greatly reduces the effects of background charge noise.

\begin{figure}
  \includegraphics[width=3in]{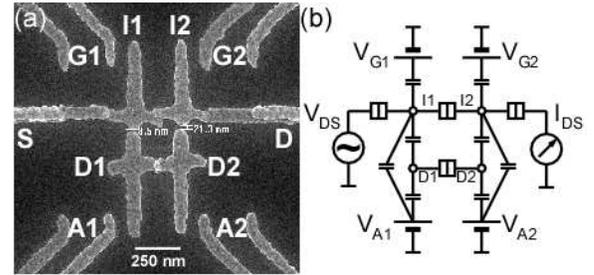}\\
  \caption{(a) SEM image of the DISET (source S, drain D and islands I1 \& I2
- connected by tunnel junctions), the capacitively coupled, floating DD (dots
D1 \& D2 connected by a tunnel junction) and four control gates (G1 \& G2 and
A1 \& A2). (b) Simplified, schematic diagram of the device showing voltage
sources and key capacitances.}\label{fig1}
\end{figure}

The devices were fabricated on a silicon substrate using standard
shadow-mask evaporation of aluminium with \textit{in-situ}
oxidation to form Al/Al$_2$O$_3$/Al tunnel junctions
\cite{fulton1987}. Fig.~\ref{fig1}a shows a SEM image of a typical
device and Fig.~\ref{fig1}b a simplified schematic. The devices
consist of a DISET, a capacitively coupled DD and four control
gates. The physical gap between the DISET and the DD was
engineered to be $< 25\:\mathrm{nm}$ in order to maximise
capacitive coupling. Electrical measurements were carried out in a
dilution refrigerator at base temperature
$T\approx20\:\mathrm{mK}$, and a magnetic field of
$B=1\:\mathrm{T}$ was applied to suppress superconductivity.
Standard, ac lock-in measurement techniques were used, with a
source-drain biase of $V_{\mathrm{DS}}=300\:\mu\mathrm{V}$
(charging energy for our DISETs
$E_{\mathrm{C}}\approx0.8\:\mathrm{m}e\mathrm{V}$).

\begin{figure}
  \includegraphics[width=3in]{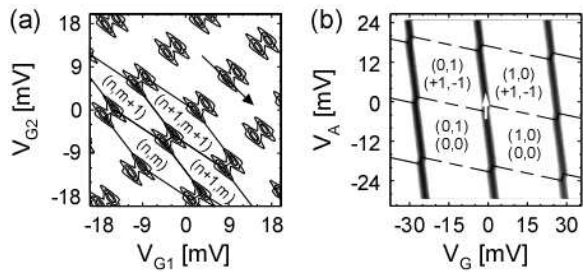}\\
  \caption{(a) Contour plot of the source-drain current $I_{\mathrm{DS}}$, exhibiting the
characteristic honeycomb pattern. Each hexagon corresponds to a specific charge
configuration $(n,m)$ of the DISET. Straight lines indicate energy degeneracy
of charge configurations, and current peaks occur where three charge
configurations are degenerate. (b) Grey-scale plot of the source-drain current
$I_{\mathrm{DS}}$ as a function of $V_{\mathrm{G}}$ and $V_{\mathrm{A}}$ (see
arrow in (a)) maps out the charge configuration space of the DISET/DD system.
The top and bottom sets of numbers denote the charge configurations of the
DISET and the DD, respectively. White corresponds to zero and black to the
maximum peak current ($181\:\mathrm{pA}$).}\label{fig2}
\end{figure}

In order to map the charge configurations of the DISET, we
recorded the source-drain current $I_{\mathrm{DS}}$ as a function
of gate voltages $V_{\mathrm{G1}}$ and $V_{\mathrm{G2}}$ (Fig.
\ref{fig2}a). Due to Coulomb blockade effects, the peaks in
$I_{\mathrm{DS}}$ define a honeycomb pattern (grey lines), the
shape of which depends on gate capacitances as well as the
inter-island coupling \cite{grabert1992}. Each hexagon corresponds
to an energetically stable charge configuration $(n,m)$ of the
DISET, where $n$ and $m$ are the number of excess electrons on
islands I1 and I2, respectively. For temperatures $T\ll
E_{\mathrm{C}}/k_{\mathrm{B}}$ and source-drain biases
$V_{\mathrm{DS}}\ll E_{\mathrm{C}}/e$, a current from source to
drain can only be observed where three charge configurations are
energy-degenerate (i.e. the system has equal electrostatic energy
for all three configurations). Such points in the charge
configuration map are termed \emph{triple-points}. Current peaks
are broadened at increased $T$ and $V_{\mathrm{DS}}$. Even at the
relatively high $V_{\mathrm{DS}}$ used in our experiments, Coulomb
blockade effects are well resolved. Since the A-gates also
capacitively couple to the DISET, a similar plot is obtained when
sweeping $V_{\mathrm{A}1}$ and $V_{\mathrm{A}2}$ (not shown),
allowing control of the charge configurations of the DD as well as
the DISET.

We now turn to study the passage of electrons from source to drain
via the DISET islands when the charge configurations of both the
DISET and the DD are electrostatically degenerate. In general, the
mechanism behind electron transport on two coupled double-dots -
irrespective of whether they are connected to source and drain
leads or not - applies to arbitrary charge configurations of the
system. However, the concept becomes clearer if one considers a
special case, where two floating double-dots are coupled in a
square arrangement with one excess electron per DD. When the
charge configurations of both double-dots are co-degenerate (i.e.
an electron has equal probability of being on either dot of a DD),
electrostatic repulsion energetically favours charge
configurations where the two electrons are diagonally opposite
each other. For symmetry reasons, the two possible configurations
are energetically degenerate, and tunnelling between the two
configurations can occur. This tunnelling primarily takes one of
two forms: energetically favourable, simultaneous tunnelling of
both electrons (co-tunnelling) \cite{averin91}, and correlated
tunnelling \cite{toth99}. The latter involves sequential
tunnelling of one electron via an excited state, followed by
tunnelling of the other electron to restore a diagonal charge
configuration. In both cases, the tunnel rate is reduced compared
to the single-electron tunnelling rate
($\Gamma=k_{\mathrm{B}}T/e^2R_{\mathrm{t}}$, where
$R_{\mathrm{t}}$ is the tunnel junction resistance)
\cite{toth99,averin91}. This reduced tunnelling rate can be
observed by replacing one of the double-dots with a DISET biased
to a triple-point. Current flow through the DISET requires the
tunnelling of single electrons between the two DISET islands. If
the charge configurations of the DISET and the DD are
co-degenerate, there is a reduction of the measured source-drain
current (compared to when the charge configurations of the DD are
\emph{non}-degenerate).

Experimental investigation of this current suppression is best
achieved by keeping the total charge of the combined system
constant, while measuring the source-drain current for different
charge configurations. The arrow in Fig.~\ref{fig2}a marks a
trajectory for G-gate biases $V_{\mathrm{G}2}=-\gamma
V_{\mathrm{G}1}$ across a triple-point, which keeps the total
charge of the DISET constant. This trajectory defines an
effective, combined G-gate bias
$V_\mathrm{G}=V_{\mathrm{G}1}\sqrt{1+\gamma^2}$. A similar
procedure was used to determine an effective A-gate bias
$V_{\mathrm{A}}=V_{\mathrm{A}1}\sqrt{1+\alpha^2}$, along which the
total charge of the DISET remains constant. Fig.~\ref{fig2}b shows
a greyscale plot of $I_{\mathrm{DS}}$ as a function of
$V_{\mathrm{A}}$ and $V_{\mathrm{G}}$. In this plot the total
charge of the combined DISET/DD system is kept constant. The upper
and lower sets of numbers denote the charge configurations of the
DISET and the DD, respectively, and define hexangular charge
configuration domains. As we are only interested in relative
changes in electron occupancy, we arbitrarily define one of the
configurations as $(0,1)(0,0)$. Due to the island-dot coupling,
the trajectories of maximum current (black) exhibit kinks each
time the charge configurations of the DISET and the DD are
co-degenerate.

\begin{figure}
  \includegraphics[width=3in]{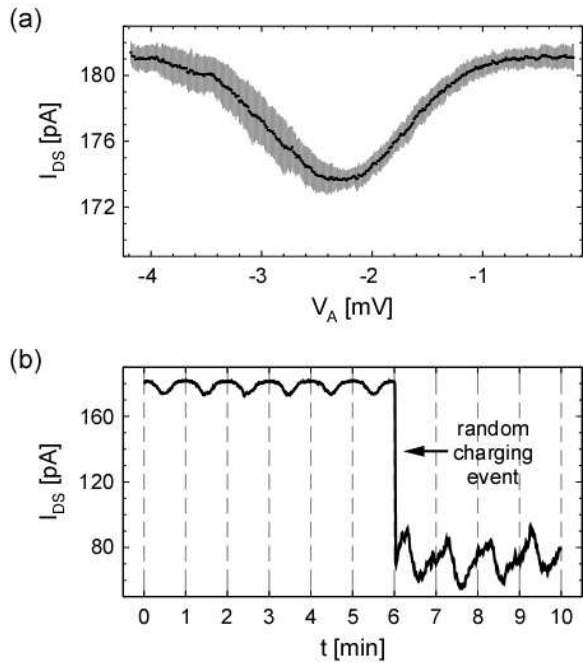}\\
  \caption{(a) A cross-section along the white arrow in Fig.~\ref{fig2}(b)
clearly shows the current suppression dip. Average and standard deviation were
obtained from 62 individual sweeps. (b) Repeated observation of the current
suppression dip shown in (a) as a function of time ($1\:\mathrm{min}$ per
sweep, followed by a fast reset). A random charging event at $t\approx
6\:\mathrm{min}$ moved the DISET off its operating point causing a reduction in
source-drain current.}\label{fig3}
\end{figure}

To study the effect degenerate charge configurations of the DD
have on the source-drain current of the DISET in more detail,
traces were taken with a constant $V_{\mathrm{G}}$ and varying
$V_{\mathrm{A}}$. $V_{\mathrm{G}}$ was chosen such that varying
$V_{\mathrm{A}}$ would include a point of co-degeneracy, and two
points where the DISET but not the DD is degenerate (arrow in
Fig.~\ref{fig2}b). Fig.~\ref{fig3}a shows the average and standard
deviation obtained from 62 individual traces. The shape of the
traces depends on temperature, source-drain bias and the
trajectory taken through gate voltage space. In this letter,
however, we focus on the absolute values of the maximum and
minimum observed source-drain current. A maximum current of
$I_{\mathrm{DS}}=181\:\mathrm{pA}$ is observable at the left and
right edge of the graph, where the DD adopts well-defined charge
configurations, and predominantly sequential single-electron
tunnelling through the DISET occurs. In the centre of the graph,
the charge configurations of the DISET and the DD are
co-degenerate. At this point, we observe a current suppression of
$\Delta I_{\mathrm{DS}}=7.5\:\mathrm{pA}$ (well above the average
noise level of $\delta I_{\mathrm{DS}}\approx0.90\:\mathrm{pA}$),
which we attribute to the reduced tunnelling rates. This behaviour
suggests that a DISET, biased to a triple-point, may be used as a
detector for degenerate charge configurations of a floating
double-dot.

We now briefly investigate the charge noise sensitivity of this
detector. SETs are known to be susceptible to random charge noise,
and it is important to discriminate such random telegraph signals
(RTSs) from signals originating from the DD. We investigated the
effect of weakly coupled charge traps, which lead to small
fluctuations of the electrostatic environment, and more strongly
coupled traps, which may induce spurious signals. To have maximum
charge sensitivity, SETs are conventionally biased to a point of
maximum transconductance ${\partial I_{\mathrm{DS}}}/{\partial
V_{\mathrm{G}}}$, which is accompanied by increased charge noise
susceptibility. The largest signal-to-noise ratios for our device,
however, were observed when biasing the DISET to triple-points,
i.e. ${\partial I_{\mathrm{DS}}}/{\partial V_{\mathrm{G}}}=0$,
where the effect of charge noise due to small fluctuations is
minimised. To investigate the influence of more strongly coupled
charge traps, we repeatedly monitored the same current suppression
feature (that shown in Fig.~\ref{fig3}a) with one minute per trace
followed by a fast reset (Fig.~\ref{fig3}b). At time
$t\approx6\:\mathrm{min}$, a large random charging event was
observed, leaving a characteristic signature: the DISET was
suddenly moved off the triple-point and the current dropped
abruptly to a lower level, where it remained for many minutes.
Empirically, RTSs that induced signals well above the average
noise level did not seem to switch back for times much longer than
typical measurement times ($\leq 1\:\mathrm{min}$). This
qualitative behaviour can be used to identify and therefore reject
such spurious signals.

So far, we have performed our experiments at low frequencies below
$500\:\mathrm{Hz}$. Fast operation, however, opens up the
possibility of investigating processes that otherwise would be
inaccessible due to fast decay or relaxation processes. In the
DISET, fast operation at radio-frequencies should be possible by
inclusion in a $LCR$ tank circuit, as has been achieved for
conventional SETs \cite{schoelkopf98} and twin-SET architectures
\cite{buehler03}.

In conclusion, we have observed current suppression due to
correlated electron transport in a DISET and a coupled, floating
DD, thereby detecting degenerate and non-degenerate charge
configurations of the DD. The maximum signal-to-noise ratio was
observed when biasing the DISET to a triple-point of its charge
configuration map. Furthermore, these biasing conditions also
provide a means for rejection of spurious charge noise.

This work was supported by the Australian Research Council, the
Australian government and by the US National Security Agency
(NSA), Advanced Research and Development Activity (ARDA) and the
Army Research Office (ARO) under contract number DAAD19-01-1-0653.
The authors would like to thank D.J. Reilly and A.J. Ferguson for
helpful discussions, and D. Barber for technical support.

\bibliography{cond-mat}

\end{document}